\documentclass[twocolumn]{aastex631}

\received{August 22, 2024}
\revised{October 31, 2024}
\accepted{October 29, 2024}

\submitjournal{ApJ}

\shorttitle{Molecular gas mass measurements of \src}
\shortauthors{Huang, Man, Lelli et al.}

\usepackage{amsmath}	
\usepackage{ragged2e}
\usepackage{symboldef}  

\begin{document}

\title[]{Molecular gas mass measurements of an active, starburst galaxy at $z\approx2.6$\\ using ALMA observations of the \CI, CO and dust emission}

\correspondingauthor{Allison W. S. Man}
\email{aman@phas.ubc.ca}

\author[0000-0003-4647-2591]{Hao-Tse Huang}
\affiliation{Department of Astronomy, University of California at Berkeley, Berkeley, CA 94720, USA}
\affiliation{Department of Physics, The Chinese University of Hong Kong, Shatin, N.T., Hong Kong}
\affiliation{Department of Physics \& Astronomy, University of British Columbia, 6224 Agricultural Road, Vancouver, BC V6T 1Z1, Canada}

\author[0000-0003-2475-124X]{Allison W. S. Man}
\affiliation{Department of Physics \& Astronomy, University of British Columbia, 6224 Agricultural Road, Vancouver, BC V6T 1Z1, Canada}

\author[0000-0002-9024-9883]{Federico Lelli}
\affiliation{INAF - Arcetri Astrophysical Observatory, Largo E. Fermi 5, 50125, Florence, Italy}

\author[0000-0002-6637-3315]{Carlos De Breuck}
\affiliation{ESO, Karl Schwarzschild strasse 2, 85748 Garching, Germany}

\author[0009-0008-8936-1625]{Laya Ghodsi}
\affiliation{Department of Physics \& Astronomy, University of British Columbia, 6224 Agricultural Road, Vancouver, BC V6T 1Z1, Canada}

\author[0000-0002-7299-2876]{Zhi-Yu Zhang}
\affiliation{School of Astronomy and Space Science, Nanjing University, Nanjing 210023, China}
\affiliation{Key Laboratory of Modern Astronomy and Astrophysics, Nanjing University, Ministry of Education, Nanjing 210023, China}

\author[0000-0002-2231-8381]{Lingrui Lin}
\affiliation{School of Astronomy and Space Science, Nanjing University, Nanjing 210023, China}
\affiliation{Key Laboratory of Modern Astronomy and Astrophysics, Nanjing University, Ministry of Education, Nanjing 210023, China}

\author[0000-0002-0818-1745]{Jing Zhou}
\affiliation{School of Astronomy and Space Science, Nanjing University, Nanjing 210023, China}
\affiliation{Key Laboratory of Modern Astronomy and Astrophysics, Nanjing University, Ministry of Education, Nanjing 210023, China}

\author[0000-0003-2733-4580]{Thomas G. Bisbas}
\affiliation{Research Center for Astronomical Computing, Zhejiang Laboratory, Hangzhou 311100, China}

\author[0000-0001-5783-6544]{Nicole P. H. Nesvadba}
\affiliation{Université de la Côte d’Azur, Observatoire de la Côte d’Azur, CNRS, Laboratoire Lagrange, Bd de l’Observatoire, CS 34229, 06304 Nice Cedex 4, France}

\begin{abstract}

We present new ALMA observations of a starburst galaxy at cosmic noon hosting a radio-loud active galactic nucleus: \src\, at $z=2.57$.
To investigate the conditions of its cold interstellar medium, we use ALMA observations which spatially resolve the \CI\ fine-structure lines, \CItwoone\, and \CIonezero, CO rotational lines, \COsevensix\, and \COfourthree, and the rest-frame continuum emission at 461 and 809\,GHz.
The four emission lines display different morphologies, suggesting spatial variation in the gas excitation conditions.
The radio jets have just broken out of the molecular gas but not through the more extended ionized gas halo.
The \CItwoone\ emission is more extended ($\approx8\,{\rm kpc}\times5\,{\rm kpc}$) than detected in previous shallower ALMA observations.
The \CI\ luminosity ratio implies an excitation temperature of $44\pm16$\,K, similar to the dust temperature.
Using the \CI\ lines, \COfourthree, and 227\,GHz dust continuum, we infer the mass of molecular gas \mmol\ using three independent approaches and typical assumptions in the literature.
All approaches point to a massive molecular gas reservoir of about $10^{11}$ \msun, but the exact values differ by up to a factor of 4.
Deep observations are critical in correctly characterizing the distribution of cold gas in high-redshift galaxies, and highlight the need to improve systematic uncertainties in inferring accurate molecular gas masses.

\end{abstract}

\section{Introduction}

Cold molecular hydrogen is the fuel for star formation. 
Investigations of the cold interstellar medium (ISM) conditions in galaxies thus inform us of the physical processes regulating their star formation activity.
Since cold molecular hydrogen (\Htwo) is not directly observable, literature studies have used tracers like carbon monoxide (CO), atomic carbon (\CI), and/or dust continuum to infer \Htwo\ mass \citep[see reviews by][and references therein]{Carilli2013,Tacconi2020,Saintonge2022}.
The calibration of these tracers of the cold ISM is not at all straightforward and remains an active research area.
The challenge lies in the abundance of these tracers relatively to \Htwo, as well as the radiative processes underlying the line or continuum emission.
For example, the CO spectral line energy distribution is a function of the gas density, temperature, and optical depth distribution \citep{Narayanan2014}, which are not easily measurable in unresolved observations of distant galaxies.
Fine-structure \CI\ emission lines follow a simpler system with only three energy levels \citep{Papadopoulos2004a}.
The \CI\ line ratios primarily depend on gas excitation temperature,
although sub-thermal excitation may also play a role under certain conditions \citep{Papadopoulos2022}.

Given the considerable uncertainties involved in \Htwo\ mass inference, it is imperative to calibrate the various methods by observing multiple \Htwo\ mass tracers for galaxies whose properties have been well-characterized. 
In this work, we perform this study using observations of CO, \CI\ and dust continuum of \src\ obtained with the Atacama Large Millimeter Array (ALMA).
\src\ (Table~\ref{tab:galaxy_info}) is one of the best studied galaxies at $z\sim2.6$ hosting an active galactic nucleus (AGN), so it provides an important test of \Htwo-mass calibrations given its complexity.
\src\ was originally identified as a high-redshift radio galaxy (HzRG) through its ultra-deep radio spectrum \citep{Roettgering1997}.
Its host galaxy is very massive with stellar mass \mstar~$=(3\pm2)\times10^{11}$\,\msun\ \citep{dBreuck2010}, and highly star-forming with star formation rate SFR = 1020\,\msun\peryr \citep{Falkendal2019}.
In parallel, its young radio jets have not yet broken out of the ionized gas outflow \citep{Nesvadba2017a}.
Deep VLT/X-Shooter spectrum revealed a bursty recent star formation history in \src. 
Shallow ALMA imaging of the \CItwoone\ emission reveals that the cold gas forms a rotating disk with a radius of $\approx4$\,kpc, which is roughly perpendicular to the more extended ionized gas outflow \citep{Lelli2018}.

In this work, we present an analysis of ALMA follow-up observations that provide detections of the \CItwoone, \CIonezero, \COsevensix, \COfourthree\ emission lines and continuum. 
The main goal is to use \src\ as a test for various \Htwo-mass calibrators and to characterize its ISM conditions.
Throughout this paper, we assume a \citet{Kroupa2001} initial mass function (IMF). A cosmology of $H_{0}$ = 70\,\kms\,Mpc$^{-1}$, $\Omega_\mathrm{M}$ = 0.3 and $\Omega_{\Lambda}$ = 0.7 is adopted.
In this cosmology, 1\arcsec\ corresponds to 8.0\,kpc at z$\sim$2.6.
We adopt the radio velocity definition.

\begin{table}
    \caption{Properties of \src}
    \begin{center}
    \begin{tabular}{l|c}
    \hline
        Property                              & Value\\
        \hline
        RA$^{(a)}$                                          & $5^h30^m25^s.44$\\
        DEC$^{(a)}$                                         & $-54\degr54\arcmin23\arcsec.19$\\
        Systemic redshift$^{(b)}$ (\zsys)               & $2.5725\pm0.0003$\\
        Stellar mass$^{(c)}$ ($10^{11}$\,\msun)             & $\simeq3\pm2$\\
        Star formation rate$^{(d)}$ (\msun$\cdot$yr$^{-1}$) & $1020^{+190}_{-170}$\\
        Molecular gas mass$^{(e)}$ ($10^{11}$\,\msun)       & $0.7-3.1$\\
        \hline
    \end{tabular}
    \end{center}
    $^{(a)}$This paper, the centroid of the fitted 2D Gaussian profile of ALMA Band 6 continuum of the host galaxy, $^{(b)}$\citet{Man2019},$^{(c)}$\citet{dBreuck2010},$^{(d)}$\citet{Falkendal2019,Man2019}, $^{(e)}$This paper - note that here the molecular gas mass is the range that includes the three values we derived from three different approaches (see Section \ref{sec:mol_mass}).
    \label{tab:galaxy_info}
\end{table}

\section{Data and Method}

\subsection{ALMA observations}\label{sec:alma_obs}

\src\ was observed with the ALMA 12\,m array using both Band 4 and Band 6 during Cycle 6 (project code 2018.1.01669.S, PI: Federico Lelli). 
The Band 4 observations were conducted in one execution block on 18 July 2019, with 43 antennae arranged in an extended configuration (C43-8) with baselines ranging from 92\,m to 8547\,m.
The mean precipitable water vapour (PWV) was $\approx$1.6-1.8\,mm. 
The total on-source integration time was 49.28\,minutes.
The four spectral windows were centered at 126.421\,GHz, 128.379\,GHz, 138.483\,GHz, and 140.129\,GHz, respectively.
The second and third spectral windows cover the \COfourthree\ and \CIonezeroLong\ (hearafter \CIonezero) transitions at the redshift of \src.
The Band 6 observations were conducted in three execution blocks on 9, 23 August, 2019 and 18 September 2019, resulting in a total on-source integration time of 81.06\,minutes.
The first execution on 9 August 2019 was taken at slightly worse weather (PWV 0.8--0.9\,mm) than the latter two executions (PWV 0.5--0.6\,mm).
The observations were conducted with 43 antennae arranged in extended configurations (roughly C43-6 and C43-7) and the baselines ranged from 15\,m to 5893\,m.
The four spectral windows were centered at 226.200\,GHz, 228.075\,GHz, 240.000\,GHz, and 241.875\,GHz, respectively, where the first spectral window covers the \CItwooneLong\ (hearafter \CItwoone) and \COsevensix\ lines.
Both Band 4 and Band 6 observations were conducted in the frequency division mode, with a native frequency resolution of 3.90625\,MHz per channel.
Each spectral window spanned a bandwidth of 1.875\,GHz and contained 480 channels each.
For observations in both bands, J0519-4546 was used as pointing, bandpass and flux calibrators, while J0550-5732 was used as phase calibrator.

Measurement sets were restored with the Common Astronomy Software Package \citep[CASA version 5.6.1-8;][]{McMullin2007} using pipeline version 42866M with the calibration script delivered by the observatory. 
The first Band 6 execution was flagged as semi-pass in the initial quality assurance (QA0) stage due to noisy check source. We calibrated the data and confirmed that the addition of the calibrated dataset leads to reduced noise in the resulting image, and thus we include all three executions in our analysis.

For image reconstruction, we used the \texttt{tclean} task in CASA version 5.8.0 to image the measurement sets.
We used \citet{Briggs1995} weighting with \texttt{robust=2} (similar to natural weighting) to optimize the signal to noise ratio (SNR) of the detection.
In Band 4 and 6, the line cubes are further binned by a factor of three and six along the frequency axis to a channel width of 11.71875\,MHz and 23.43704\,MHz, respectively ($\approx$ 26\,\kms\ and 31\,\kms\ near the \CIonezero\ and \CItwoone\ emission lines).
The continuum images are created using line-free channels.
For Band 6, the entire spectral window containing the \CItwoone\ and \COsevensix\ lines are excluded, and a continuum image is made using data from the remaining three spectral windows.
For Band 4, the following frequencies are identified to be far from emission lines based on an inspection of the cube, and are used to create the continuum images: 126.031--127.332\,GHz, 127.543--128.9\,GHz, 138.2--139.2\,GHz, and 139.267--140.743\,GHz.
The synthesized beams and resulting root-mean-square (RMS) noise of the reconstructed images are listed in Table~\ref{tab:alma_obs_log}.
Note that we do not include flux calibration error in our analysis. The scale of flux calibration error is estimated to be around 5 percent \citep{ALMA2021}.

Shallower ALMA Band 6 observations of \src\ were analyzed and published in \citet{Lelli2018} and \citet{Man2019}, using Cycle 2 observations obtained via project code 2013.1.00521.S (PI: Carlos De Breuck).
The Cycle 2 observations had a slightly different spectral setup, such that the \COsevensix\ line was too near the edge of the bandpass to be useful for analysis. Due to the shallow depth of the Cycle 2 data (integration time of 5\,minutes) and the different spectral set up, no attempt was made to combine the Cycle 2 data with the Cycle 6 data for the image reconstruction.

\begin{deluxetable*}{ccccccc}
    \tablewidth{0pt} 
    \tablecaption{ALMA Observation Log (Program ID: 2018.1.01669.S) \label{tab:alma_obs_log}}
    \tablehead{
    \colhead{Band} & \colhead{Date} & \colhead{ToS} & \colhead{Baseline Range} & \colhead{Synthesized Beam$\dagger$} & \colhead{Continuum RMS} & \colhead{Line RMS}\\
    \colhead{} & \colhead{} & \colhead{(s)} & \colhead{(m)} & \colhead{} & \colhead{(\uJy\,\perbeam)} & \colhead{(mJy\,\perbeam)}
    }
    \startdata 
        4 & 18-Jul-2019 & 5565 & 92.1 - 8547.6 & 0.13\arcsec$\times$0.12\arcsec\ at PA 48\degree\ & 10 & $\approx0.25$ per 25.5\,\kms\ channel\\
        6 & 23-Aug-2019, 18-Sep-2019 & 4904 & 15.1 - 3396.5 & 0.14\arcsec$\times$0.13\arcsec at PA $-46$\degree\ & 11 & $\approx0.17$ per 31.0\,\kms\ channel \\
    \enddata
    \vspace{2mm}
    \justifying
    \noindent
    $\dagger$The quoted native synthesized beam is only for continuum images. The line cubes and moment-zero maps (Figure \ref{fig:moment0_overview}) were made using a restoring beam of size $0.2\arcsec\times0.2\arcsec$.
    The line RMS is calculated from the data cube prior to the continuum subtraction and is the average of all the channels in the spectral window that contains \CI\ line.
    Both the continuum and line RMS are estimated from an annulus centered on the host galaxy, with inner and outer diameters of 3$\arcsec$ and 10$\arcsec$.
\end{deluxetable*}

\subsection{Continuum modeling and subtraction}
\label{sec:contsub}

The Cycle 6 observations have sufficient spatial resolution to resolve the continuum emission of the host galaxy from the eastern radio lobe. Given the spatially and spectrally varying nature of the continuum, we have attempted several ways to model the continuum and subtract it for the subsequent spectral line analysis.

Generally, continuum subtraction performed on the uv-plane becomes less reliable the further away from the phase center\footnote{\url{https://casadocs.readthedocs.io/en/stable/api/tt/casatasks.manipulation.uvcontsub.html}}.
We compared the results of continuum subtraction on the uv-plane and the image plane using the \texttt{uvcontsub} and the \texttt{imcontsub} tasks in \texttt{CASA}, respectively.
We found that the continuum appeared to be over-subtracted when performed on the uv-plane especially in Band 6.
We thus subtracted the continuum on the image plane using \texttt{imcontsub} by modeling it as a zeroth-order polynomial across each side band (i.e., two adjacent spectral windows) containing a given line.
The continuum is fitted to line-free channels that exclude noisy ones (e.g., edge channels).
The continuum-subtracted cubes are used for creating the moment maps.

\subsection{Spectrum extraction}
\label{sec:spec}
\begin{figure*}
    \includegraphics[width=\textwidth]{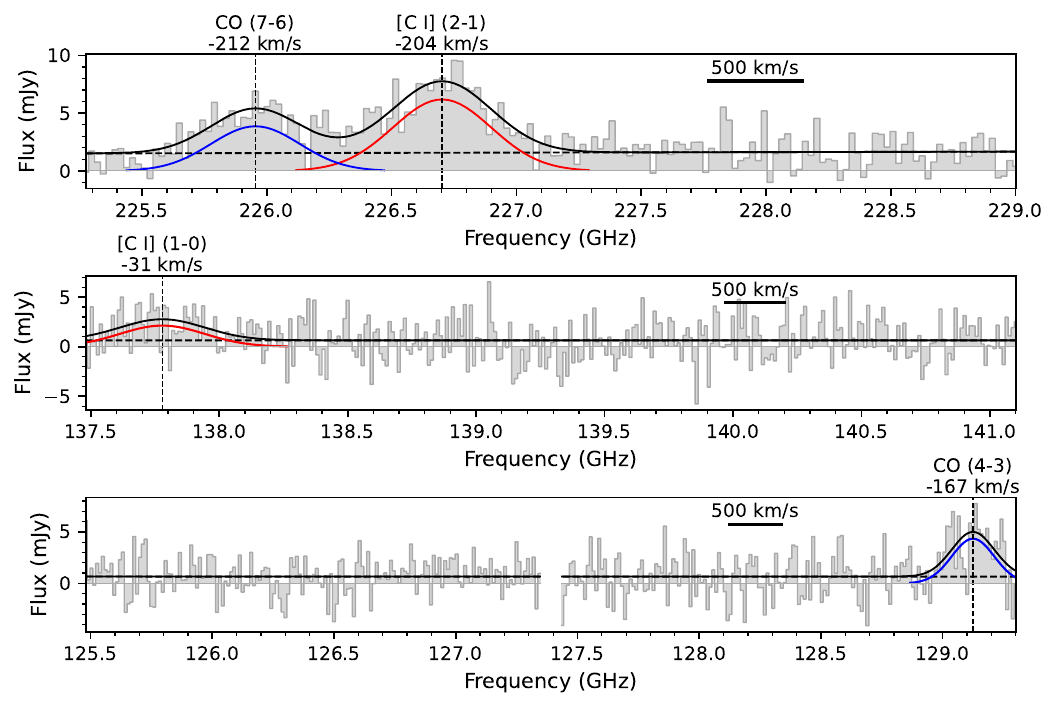}
    \caption{The extracted ALMA spectrum of \src.
    The top panel shows the Band 6 spectrum and the middle and bottom panels show the Band 4 spectrum.
    On top of the observed spectrum (shaded in gray), we overlay the best-fitted model using a black solid line.
    The model consists of assigning a Gaussian profile to each emission line and a 1D polynomial to the underlying continuum.
    The component of the continuum is plotted as a horizontal, black-dashed line and the components of emission lines are plotted in red and blue lines for \CI\ and CO, respectively.
    The emission lines and the velocity centers of the fitted model, calculated with respect to \zsys= 2.5725, are labeled at the top of each panel.
    An additional scale bar is provided to convert the frequency scale to the velocity scale.
    The channel widths are 11.72\,MHz and 23.44\,MHz for the Band 4 and Band 6 spectrum, respectively.
    While we only display one sideband of the spectrum in each panel, the fitting was performed by making use of both sidebands to maximize the number of channels for the continuum (see Section \ref{sec:line_analysis}).
    Note that the continuum levels presented here do not properly account for the separation of different emission components and should not be used in further analysis (see Section \ref{sec:cont_fit} for the analysis of the spatially resolved continuum emission).}
    \label{fig:spectrum_overview}
\end{figure*}

The emission lines detected in the ALMA observations include \CItwoone, \COsevensix\ (Band 6), and \CIonezero, \COfourthree\ (Band 4).
To ensure consistency, we used the same aperture to extract the spectrum shown in Figure \ref{fig:spectrum_overview}.
First, a $2\arcsec\times2\arcsec$ square aperture centered on the host galaxy was used to extract the preliminary spectrum and identify the rough velocity ranges of the emission lines.
Tentative moment-zero maps were then constructed from those velocity ranges for the inspection of the spatial distribution of the emission lines.
We found that both \CItwoone\ and \COsevensix\ showed an extended structure towards the South-West direction (see Figure~\ref{fig:moment0_overview} and Section~\ref{sec:line_morphology}), and therefore chose an inclined elliptical aperture to include the spatially extended emission for spectral extraction.
Such extended emissions were not detected in previous shallower ALMA observations and thus a smaller aperture was used in \citet{Man2019}.
The extracted spectrum is presented in Figure \ref{fig:spectrum_overview}. A detailed description of emission line fitting is presented in Section \ref{sec:spectrum_analysis}.  The properties of the emission lines are listed in Table \ref{tab:line_info}.

\section{Results}

\subsection{Continuum fitting}
\label{sec:cont_fit}
\begin{figure*}
	\includegraphics[width=\textwidth]{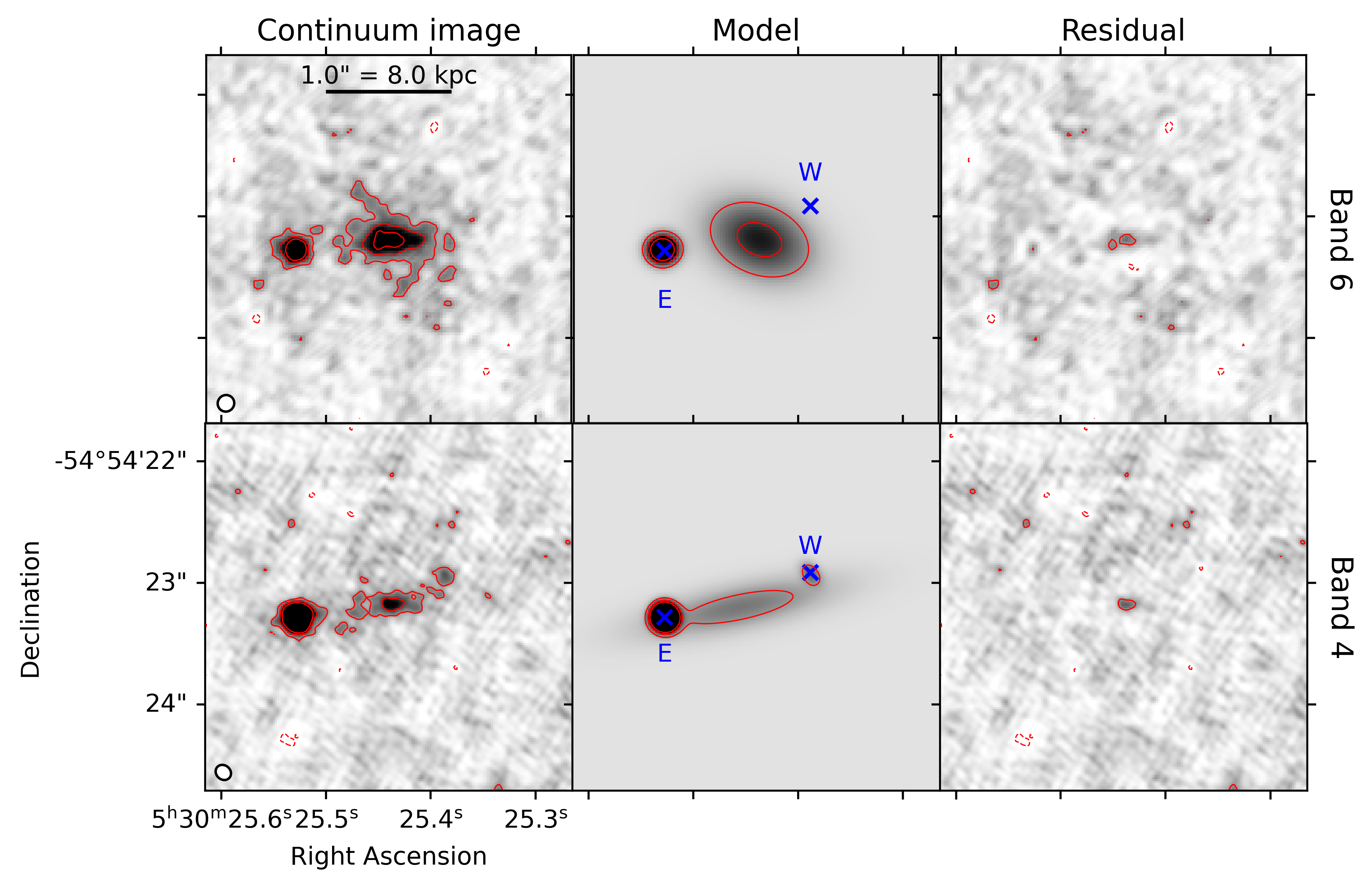}
    \caption{The best-fit models of ALMA Band 4 (bottom row) and Band 6 (top row) continuum images using \texttt{imfit} from CASA.
    In each row, the left panel is the continuum image, the middle panel is the best-fit model, and the right panel is the residual image.
    The red contours have levels [-3, 3, 6, 9]\,$\sigma$, where $\sigma$\ = 10.6\,\uJy\,\perbeam and 10.1\,\uJy\,\perbeam\ for Band 4 and Band 6, respectively.
    In the middle panels (model images), the positions of the East and West radio lobes are marked with blue crosses.
    The positions of the radio lobes are estimated from the 2D Gaussian peaks of the Band 4 continuum fitting.
    The synthesized beam for each band is plotted at the bottom left corner of the left panel.
    The angular size of each image is 3\arcsec.
    Due to the irregular shape of the continuum of the host galaxy, the 2D Gaussian cannot fully capture the host galaxy emission and there are residual emissions in the center in both bands (see details in Section \ref{sec:cont_fit}).
    This reflects the limitation of \texttt{imfit}, and the potential blending of different components.}
    \label{fig:cont_fit}
\end{figure*}

The ALMA Band 4 and 6 continuum images are presented in Figure \ref{fig:cont_fit} and \ref{fig:moment0_overview}.
In both continuum images, a compact emission on the left (East) and a more diffuse emission at the center can be seen.
The nearly identical position of the Eastern, compact emission with the Eastern radio lobe detected in the 18.5 GHz continuum with ATCA \citep{Broderick2007} suggests that this emission is the continuation of the power law of synchrotron emission at higher frequencies \citep[see also][]{Falkendal2019}.
The central, more diffuse emission is co-spatial with the \CItwoone\ emission, indicating that this emission originates from the dust thermal emission within the host galaxy.
For the Band 4 continuum image, in addition to these two emission regions, a fainter, compact emission is further detected in the North-West direction relative to the central, diffuse emission ($\approx0.65\arcsec$\ apart).
The position of this faint emission coincides with the Western radio lobe detected in the ATCA 18.5 GHz continuum, again suggesting that it originates from synchrotron emission.

From the above comparison,
we identified two synchrotron emission components and one dust thermal emission component in the Band 4 continuum image, and one synchrotron emission component and one dust thermal emission component in the Band 6 continuum image.
We measure the fluxes of these different components in the ALMA continuum images by using the CASA task \texttt{imfit} to fit a 2D Gaussian profile to each component.
The best-fit models and their residuals are presented in Figure \ref{fig:cont_fit}.
The details of each component are shown in Table \ref{tab:continuum_info}.

For the ALMA Band 6 continuum, the fitted flux of synchrotron emission agrees with the best-fit model presented in \citet{Falkendal2019} within errors, while the determined flux of dust emission ($1.55\pm0.14$\ mJy) is slightly higher than the value presented in \citet{Falkendal2019} ($1.33\pm0.16$\ mJy), though still consistent within 1.4\,$\sigma$.
On the other hand, the best-fit model for ALMA Band 4 continuum exhibits an unusually elongated central component compared to the ALMA Band 6 counterpart, in addition to having a $\sim34\degr$\ difference in orientation, suggesting that the fitting routine struggled to deblend the continuum emission from the host galaxy from the fainter, western radio lobe.
This may be caused by the limitation of \texttt{imfit}, and/or a more complex and blended Band 4 continuum morphology due to the stronger synchrotron emission and weaker dust continuum emission relative to Band 6.
A hidden synchrotron emission from the radio jet may also be present in the host galaxy, as hinted by the elongated morphology of the Band 4 continuum along the direction between the two radio lobes (see the $3\sigma$ contour of the continuum in Figure \ref{fig:cont_fit}).
Future high-resolution radio observations will be needed to confirm this scenario.

To determine the validity of dust emission flux obtained by \texttt{imfit} in ALMA Band 4, we computed the slope of the dust emission at ALMA Band 4 and Band 6 by modeling the measured fluxes as a power law ($\nu^\alpha$).
The obtained spectral index is $\alpha\,=\,0.63\,\pm\,0.29$.
This value is much lower than the value used in \citet{Falkendal2019} for their spectral energy distribution (SED) fitting ($\alpha$=4.5), and even smaller than what we would expect if the dust is optically thick \citep[$\alpha\,=\,2$,][]{Rybicki1986}.
At the wavelength observed by ALMA Band 6 and Band 4, the dust thermal emission is usually optically thin, with $\alpha=3-4$\ \citep{Scoville2016}.
The low spectral index obtained from \texttt{imfit} may thus result from the unaccounted flux from synchrotron emission in the host galaxy.
Due to the irregular shape of the ALMA Band 4 continuum and the unclear nature of the low spectral index, we present the fitted fluxes of ALMA Band 4 continuum in Table \ref{tab:continuum_info} but exclude them from subsequent analysis.

\subsection{Emission line analysis}
\label{sec:line_analysis}
\subsubsection{Spectral analysis}
\label{sec:spectrum_analysis}
The emission lines detected in the ALMA observations include \CItwoone, \COsevensix\ from Band 6 and \CIonezero, \COfourthree\ from Band 4.
Since the continuum emission was detected in both bands, the analysis of emission line fluxes must be done simultaneously with the estimation of the continuum level.
We model each emission line with a Gaussian profile and the continuum as a first-order polynomial\footnote{We find that modeling continuum on the 1D spectrum yields more reliable results than during the imaging (Section~\ref{sec:cont_fit}), due to the complex morphology of the continuum emission.}.
The extracted spectra and the best-fit models are plotted in Figure \ref{fig:spectrum_overview}.

To estimate the uncertainty in the fitted emission line profiles, we utilized the Monte Carlo technique to re-sample and fit the spectrum 1000 times.
During each realization, we first constructed a new spectrum by adding Gaussian noise independently to each channel of the observed spectrum.
The uncertainty in the integrated flux per velocity channel, $\sigma_{\text{int}}$, was estimated via
\begin{equation}
    \sigma_{\text{int}}=\sqrt{N}\cdot\sigma_{\text{rms}},
\end{equation}
where $N$ is the ratio of the size of the extraction aperture to that of the synthesized beam and $\sigma_{\text{rms}}$ is the rms error per velocity channel per pixel.
The rough values of $\sigma_{\text{rms}}$ are listed in Table \ref{tab:alma_obs_log}.
For each realization, we fit the spectrum with the Gaussian components and the continuum as described above.
The means and standard deviations of the parameters of the fitted line profiles from 1000 Monte Carlo realizations are presented in Table \ref{tab:line_info}.

\subsubsection{Emission line morphology}\label{sec:line_morphology}
\begin{figure*}
	\includegraphics[width=\textwidth]{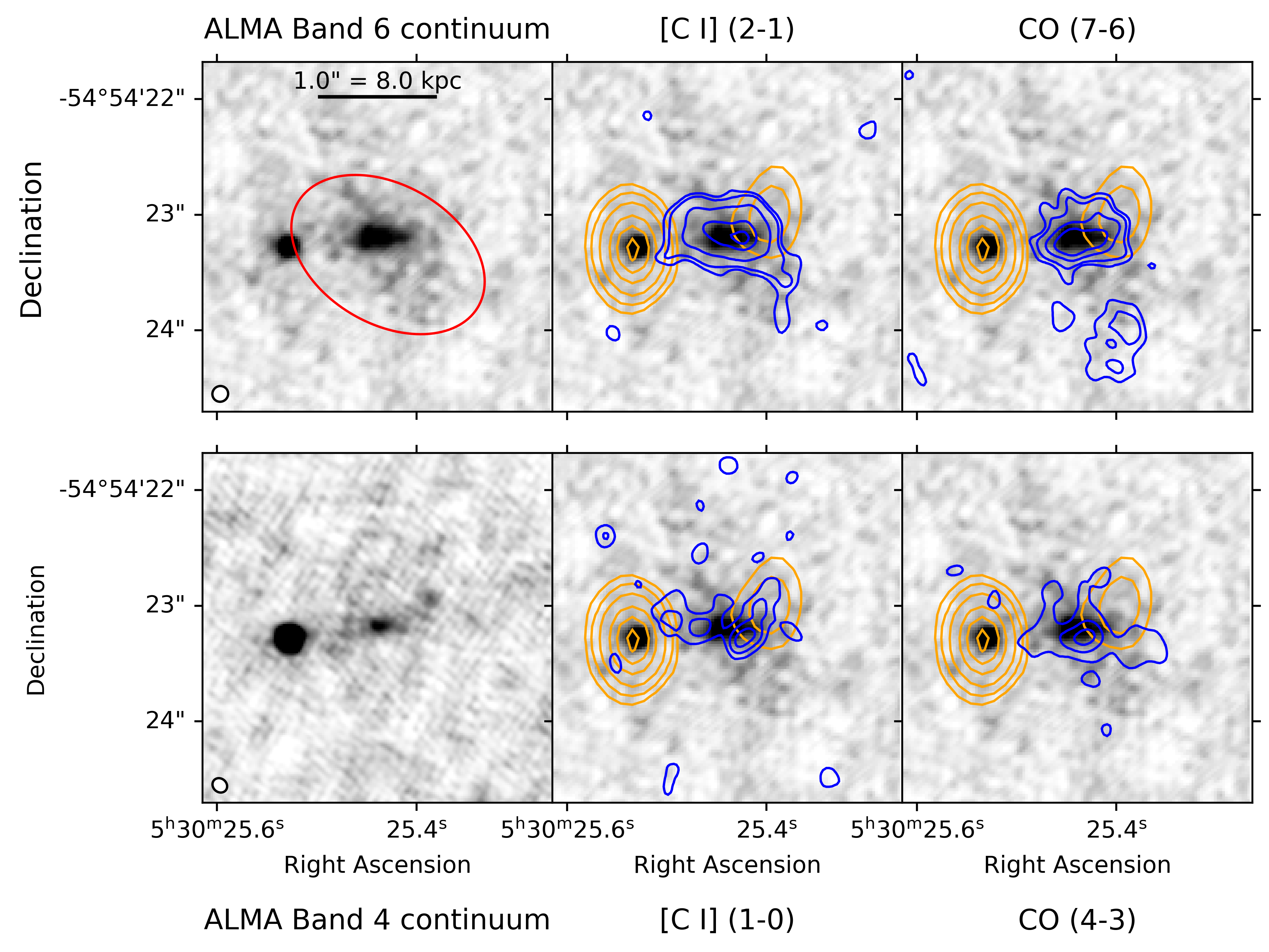}
    \caption{The continuum maps of ALMA Band 4, 6 and moment-zero maps of four emission lines of \src.
    The top left and bottom left panels are the grayscale images of the ALMA Band 6 and Band 4 continuum, respectively.
    The red ellipse in the top left panel is the spectrum extraction aperture in this paper.
    The synthesized beams of the ALMA Band 4, 6 continuum images are shown as ellipses in the bottom left corners of the left panels.
    For the remaining four panels, we used the ALMA Band 6 continuum in the top left panel as the background image, and overlaid the contours of the moment-zero maps of the corresponding lines (blue) and the ATCA 18.5\,GHz radio continuum (orange).
    Note that the line cubes and moment-zero maps are produced with a uniform beam size of $0.2\arcsec\times0.2\arcsec$.
    For ATCA 18.5\,GHz radio continuum, the contour levels are 1, 2, 4, 8, 12, 16 mJy\,\perbeam.
    Due to the difference in the depths of ALMA Band 4 and Band 6 observations and the intrinsic strengths of the lines, the blue contours in the panels of emission lines do not represent the same intensity level.
    The contour levels are [2, 4, 6, 8, 12, 16]\,$\sigma$\ for \CItwoone, with $\sigma$\ = 37\,mJy\,\perbeam$\cdot$\kms; [2, 4, 6, 8]\,$\sigma$\ for \COsevensix, with $\sigma$\ = 38\,mJy\,\perbeam$\cdot$\kms; [2, 3, 4]\,$\sigma$\ for \CIonezero, with $\sigma$\ = 75\,mJy\,\perbeam$\cdot$\kms; [2, 3, 4, 6]\,$\sigma$\ for \COfourthree, with $\sigma$\ = 55\,mJy\,\perbeam$\cdot$\kms, respectively.}
    \label{fig:moment0_overview}
\end{figure*}

We constructed moment-zero maps for the emission lines using the \texttt{moment} task from Python package \texttt{spectral-cube}\footnote{\url{https://spectral-cube.readthedocs.io/en/latest/moments.html\#moment-maps}}.
The integration range of velocity was chosen to be $[v_{c}-2\sigma_v,v_{c}+2\sigma_v]$, where $v_{c}$ and $\sigma_v$ are the center of velocity and standard deviation
of the best-fit Gaussian profile for each emission line as listed in Table \ref{tab:line_info}.
The contours of the moment-zero maps using the continuum-subtracted data are presented in Figure \ref{fig:moment0_overview}.

Out of the four detected emission lines, \CItwoone\ displays the most significant detection, with the East and West edges of emission approaching the centers of synchrotron emission.
We used the emission of \CItwoone\ to estimate the projected spatial extent of molecular gas.
Using the 3$\sigma$ contour of the \CItwoone\ moment-zero map, the size of emitting region is estimated to be  (7.7\,$\pm$\,0.9)$\times$(4.9\,$\pm$\,0.9) kpc
\footnote{The corresponding angular size is (0.96\arcsec\,$\pm$\,0.11\arcsec)$\times$(0.61\arcsec\,$\pm$\,0.11\arcsec).
Due to the irregular shape of \CItwoone\ emission, the direction and the size of the disk were measured by visual inspection.
The measurement was done on beam-convolved data without correcting for beam smearing.},
with a position angle of $\sim$90\degr\ from North to East.
The error in angular size is estimated as $\sqrt{2}\cdot\mathrm{beam_{FWHP}/SNR}/0.9$, where $\mathrm{beam_{FWHP}}$ is the Full-Width-Half-Power synthesized beam size, $\mathrm{SNR}=3$ is the signal-to-noise ratio of the contour, and an additional factor of $\sqrt2$ takes into account that measurements of two sides are needed to obtain the length \citep{ALMA2021}.
Our measurement is consistent with the results of \citet{Lelli2018}, who estimated a \CItwoone\ size between 4 and 8 kpc and a position angle of $75\pm12$\degr.
The emission of \COsevensix\ is more compact than \CItwoone, which is expected since the emission of high-J CO transitions requires molecular gas of high temperature and density. 

The emission of \CIonezero\ and \COfourthree\ exhibits more irregular morphology compared to \CItwoone\ and \COsevensix, which may partially be the result of the lower SNR of the Band 4 data.
Nonetheless, even if we focus exclusively on the central region of the galaxy, where the detection is the most significant, we still see a noticeable difference in the morphology of \CItwoone\ and \CIonezero.
Contrary to a singly peaked emission of \CItwoone, the emission of \CIonezero\ appears to be doubly peaked.
The nature of this difference is unclear at this moment; however, it has the potential to provide a unique way to probe the \CI\ excitation mechanisms at the center of the galaxy, using the correlation of \CI\ emission ratio to the kinematic temperature of the molecular gas \citep[see Section \ref{sec:mol_mass_CI} and][]{Papadopoulos2004a,Papadopoulos2004b,Schneider2003}.

Another interesting observation of the emission morphology is the extended line emission towards the southwest direction seen in \CItwoone\ and \COsevensix, whose extent is $\sim$7\,kpc measured from the galactic center.
The extended emission looks like a tail in \CItwoone\ and there appears to be an isolated blob at a further south position in \COsevensix.
This extended emission does not appear in \CIonezero\ and \COfourthree.
The reason for this absence is unclear:
it is likely that this extended emission may have been missed due to the lower SNR in the ALMA Band 4 observations.
The origin of the blob and tail seen in \COsevensix\ and \CItwoone\ is uncertain.
Their direction is different from both the major axis of \CIonezero\ and the radio lobes, so these features are probably unrelated to the ionized gas outflow seen in SINFONI data (see \citealt{Lelli2018}).
In a companion paper (Lin et al., accepted in \aap), we present a detailed 3D kinematic analysis of the \CItwoone\, data, which reveals a complex situation with multiple kinematic components: (1) an inner rotation-supported disk \citep[as in][]{Lelli2018}, (2) a second, weaker tail to the East that is projected on top of the rotating disk, and (3) a non-circular component at $\sim$2 kpc to the South of the galaxy center. A possible interpretation is that the two tails are the ``leftover'' tidal tails of a past major merger event, which may have triggered a gas inflow as well as the starburst and AGN activity.
Future deeper observations will be required to properly identify the nature of this galaxy.

\begin{table*}
	\caption{Emission line measurements}
	\label{tab:line_info}
        \begin{center}
	\begin{tabular}{lcccc}
		\hline
		Line            &  $v-v_{sys}$   & FWHM          & ${S\Delta v}$     & $L^{'}$\\
		                  & (\kms)        & (\kms)        & (Jy$\cdot$\kms)   & ($10^{10}$\,K$\cdot$\kms$\cdot$pc$^2$)\\
		\hline
		$\CItwoone$     & -204 $\pm$ 17 & 615 $\pm$ 56 & 4.04 $\pm$ 0.30 & 2.50 $\pm$ 0.19\\
		$\COsevensix$   & -212 $\pm$ 31 & 555 $\pm$ 71 & 2.27 $\pm$ 0.27 & 1.42 $\pm$ 0.17\\
		$\CIonezero$    & -31 $\pm$ 82 & 848 $\pm$ 230 & 1.86 $\pm$ 0.40 & 3.12 $\pm$ 0.67\\
		$\COfourthree$  & -167 $\pm$ 31 & 476 $\pm$ 47 & 2.19 $\pm$ 0.26 & 4.18 $\pm$ 0.50\\
		\hline
	\end{tabular}
        \end{center}
	\justifying
        Table Note: 
        Each column (from left to right) represents the emission line velocity center with respect to the systematic redshift $z=2.5725$, Full-Width-Half-Maximum (FWHM), integrated flux, and line luminosity.
	The values and errors for velocity center, FWHM, and integrated flux are derived using the Monte Carlo technique (see Sec.~\ref{sec:spectrum_analysis}).
\end{table*}

\begin{table*}
	\caption{Continuum measurements.}
	\label{tab:continuum_info}
    \begin{center}
    	\begin{tabular}{lllc}
    		\hline
                Band             & Component      & Property             & Value \\
                \hline
                4 (129.053\,GHz) & Radio lobe (E) & flux (mJy)           & 0.751 $\pm$ 0.023 \\
                                 & Radio lobe (W) & flux (mJy)           & 0.048 $\pm$ 0.021 \\
                                 & Host galaxy    & flux (mJy)           & 1.09 $\pm$ 0.13 \\
                                 &                & Major axis (\arcsec) & 1.35 $\pm$ 0.18 \\
                                 &                & Minor axis (\arcsec) & 0.31 $\pm$ 0.04 \\
                                 &                & Position angle (\degr)       & 102 $\pm$ 3 \\
                6 (226.548\,GHz) & Radio lobe (E) & flux (mJy)           & 0.370 $\pm$ 0.034 \\
                                 & Radio lobe (W) & flux (mJy)           & $\ldots$ \\
                                 & Host galaxy    & flux (mJy)           & 1.55 $\pm$ 0.13 \\
                                 &                & Major axis (\arcsec) & 0.75 $\pm$ 0.07 \\
                                 &                & Minor axis (\arcsec) & 0.51 $\pm$ 0.04 \\
                                 &                & Position angle (\degr)       & 68 $\pm$ 9 \\
    		\hline
    	\end{tabular}
    \end{center}
    Table Note: The fitting is done by CASA \texttt{imfit}. The fitted models are plotted in Figure \ref{fig:cont_fit}.
    The value of major and minor axes for the host Galaxy are convolved with the synthesized beam. 
	In Band 6 we did not detect the synchrotron emission from East radio lobe.
	Note that the values for ALMA Band 4 are unreliable as explained in Section \ref{sec:cont_fit}, and should not be used in further analysis.
 
\end{table*}

\subsection{Inference of molecular gas masses}
\label{sec:mol_mass}
In light of the detection of multiple emission lines, along with the dust continuum, we are able to derive the molecular gas mass via three different independent approaches. 
The inferred molecular gas masses are summarized in Table \ref{tab:mmol_mass}.
The three approaches are based on several different assumptions and the results are in tension (see Section \ref{sec:mol_mass_sys_error}).

\subsubsection{The Atomic Carbon approach}
\label{sec:mol_mass_CI}
The emission lines of atomic carbon, \CItwoone\ and \CIonezero, are tracers of molecular gas \citep{Papadopoulos2004a,Papadopoulos2004b} in galaxies \citep[see][for a compilation]{Valentino2020}.
The derivation of molecular gas mass from atomic carbon requires first the estimation of atomic carbon mass, for which the excitation temperature ($\Tex$) must be used to properly constrain the population at the ground state.
The detection of both atomic carbon fine-structure transition lines, \CItwoone\ and \CIonezero, allows the determination of $\Tex$ under the assumption of local thermodynamic equilibrium (LTE).  
As derived in the Appendix of \citet{Schneider2003}, by assuming that both of the two \CI\ lines are optically thin, $\Tex$\ can be obtained using
\begin{equation}
    \Tex\,[\text{K}]=\frac{38.8}{\text{ln}(2.11/R)} \label{eqt:T_ex}
\end{equation}
where $R$ is the ratio between \CItwoone\ and \CIonezero\ line luminosities $L^{'}$ in units of K$\cdot$\kms$\cdot$pc$^2$.
We used equation (3) in \citet{Solomon1992} to convert the observed line fluxes into line luminosities and derived $R\,=\,0.84\,\pm\,0.20$ and $\Tex=44\,\pm\,16$\,K, where the errors indicate the uncertainty from line flux measurements.

Following \citet{Weiss2003}, we obtained the atomic carbon mass \mCI\ $=\,(4.1\,\pm\,0.8)\times10^{7}\,\text{\msun}$\ from the line luminosity of \CItwoone\ and $\Tex$ using
\begin{equation}
    \text{\mCI}\,[\text{\msun}]=4.566\times10^{-4}Q(\Tex)\frac{1}{5}e^{62.5/\Tex}L^{'}_{\text{\CItwoone}}
\end{equation}
where $Q(\Tex)\,=\,1+3e^{-23.6/\Tex}+5e^{-62.5/\Tex}$ is the partition function of atomic carbon.\footnote{In the literature, several definitions of $Q$ exist. Apart from $Q(\Tex)$, the factors $Q_{21}$ and $Q_{10}$ denote the number ratios of \mbox{[C\,{\sc i}]\,$^3$P$_{2}$} and \mbox{[C\,{\sc i}]\,$^3$P$_{1}$} to the entire \CI\ population, respectively \citep{Papadopoulos2004a}.\label{footnote:Q}} To get the molecular gas mass \mmol, we adapted the atomic carbon abundance \XCI$\,=\,3\times10^{-5}$ as derived by \citet{Weiss2003} for M82, and included the correction factor 1.36 for the Helium and heavier elements. This gives \mmol$\,=\,(3.1\pm0.6)\times10^{11}\text{\msun}$.
This value is nearly an order of magnitude higher than that published in \citet{Man2019}, which is based on shallower ALMA observations of the \CItwoone\ emission line.
In Appendix~\ref{sec:CI_comparison}, we provide a detailed description of the factors contributing to this discrepancy.

Some recent studies \citep{Papadopoulos2022,Dunne2022} have raised the concern that the LTE assumption in this \CI\ approach is inadequate and that the \CItwoone\ alone is not a good tracer for the molecular gas mass due to the large variation in $Q_{21}$.
Those studies suggest that using \CIonezero\ and a conversion factor $Q_{10}=0.48\,\pm\,0.08$ instead provides a more robust estimation of \mmol.
As a check of the validity of our approach, we compute the corresponding $Q_{10}$ factor from our \CI\ lines and the LTE assumption.
Our value, $Q_{10}=0.44\,\pm\,0.02$, is 8 percent lower than \citet{Papadopoulos2022}'s suggested value.
Therefore, our estimated value of \mmol\ is unlikely to be significantly impacted by our LTE assumption.
Nonetheless, the \CI\ approach does suffer from systematic errors, as discussed in Section \ref{sec:mol_mass_sys_error}.

\subsubsection{The CO approach}
The rotational lines of CO have been used extensively to probe the condition of the ISM and the mass of molecular gas \citep{Bolatto2013}.
This estimation of mass relies on the CO-to-H$_{2}$ factor, $\alphaCO\equiv$\ \mmol$/\LineLum{\COonezero}$, to convert the line luminosity of \COonezero\ to molecular gas mass\footnote{Note that this $\alphaCO$ factor already includes the correction factor 1.36 for Helium mass \citep{Bolatto2013}.}.
Since we do not have observations of the \COonezero\ line, an extra step must be taken to convert $\LineLum{\COfourthree}$ to $\LineLum{\COonezero}.$\footnote{While we have the measured $\LineLum{\COsevensix}$, we chose not to use it to estimate $\LineLum{\COonezero}$ as the higher transitions of CO emission require denser and/or hotter gas and are less representative of the bulk of molecular gas, and the conversion is thus less certain.}
The exact value for this conversion varies across galaxy types and depends on the ISM conditions.
Here we adopt the average value in the literature derived for sub-millimeter galaxies (SMGs) similar to \src, i.e.  $r_{41}\equiv\LineLum{\COfourthree}/\LineLum{\COonezero}=0.46$ \citep{Carilli2013}.
This gives $\LineLum{\COonezero}=(9.1\pm1.1)\times10^{10}$\,K$\cdot$\kms$\cdot$pc$^2$.
For $\alphaCO$, we adopted the value commonly used for starburst galaxies, 0.8\,\msun/(K$\cdot$\kms$\cdot$pc$^2$) \citep{Bolatto2013}. The final derived \mmol\ is $(7.3\ \pm\ 0.9)\times10^{10}$\msun, which is a factor of $\approx$4 smaller than the mass obtained from the \CI\ lines.

\subsubsection{The dust emission approach}
The dust continuum in the far-infrared/sub-millimeter range can provide an alternative way of estimating the molecular gas mass \citep{Scoville2016}.
The frequency range covered by ALMA Band 4 and Band 6 corresponds to the Rayleigh-Jeans tail of the modified black-body emission for typical ISM dust temperatures (a few tens of K), which is usually optically thin.
The optically thin property provides a linear scaling relation of molecular gas mass \mmol\ and dust emission luminosity $L_{\text{dust}}$ at a fixed frequency, assuming a fixed dust-to-gas abundance ratio.
\citet{Scoville2016} used a sample of 72 galaxies, including local star-forming galaxies, low-z ultraluminous infrared galaxies (ULIRGs), and SMGs, to calibrate the dust-to-gas ratio $L_{\text{dust,850\um}}/\text{\mmol}=\alphadust$ and found a remarkably similar value across different galaxy types.
Here we use $\alphadust=6.7\cdot10^{19}$\,erg\,s$^{-1}$\,Hz$^{-1}$\,\msun$^{-1}$ and equation (16) in \citet{Scoville2016} to derive the molecular gas mass:
\label{eq:mmol_dust}\begin{align}
M_{\mathrm{mol}}=& 1.78\,S_{\nu_{\text{obs}}}\,[\text{mJy}]\ (1+z)^{-4.8} \nonumber\\
              & \times\left(\frac{\nu_{850\text{\um}}}{\nu_{\text{obs}}}\right)^{3.8}\ (D_{L}\,[\text{Gpc}])^{2} \nonumber\\
              & \times\left(\frac{6.7\times10^{19}\,\text{erg\,s$^{-1}$\,Hz$^{-1}$\,\msun$^{-1}$}}{\alphadust}\right)\ \frac{\Gamma_{0}}{\Gamma_{\text{RJ}}}\ 10^{10}\text{\msun}
\end{align}
where $\Gamma_{0}=0.71$ and $\Gamma_{\text{RJ}}$ is the ratio of blackbody radiation and Rayleigh-Jeans tail assuming a mass-weighted dust temperature of $\Td=25$\,K.
As mentioned in Sec.~\ref{sec:cont_fit}, the Band 4 dust continuum emission is potentially contaminated by the synchrotron component.
We thus derived the molecular gas mass from the Band 6 dust continuum to be \mmol\ = (2.2 $\pm$ 0.2)$\times10^{11}$\msun.

\begin{table}
    \caption{Inferred molecular gas masses.}
    \label{tab:mmol_mass}
    \begin{center}
    \begin{tabular}{lc}
        \hline
        Method                  & \mmol \\
                                & ($10^{11}$\,\msun) \\
        \hline
        \CI                     & (3.1  $\pm$ 0.6)$\displaystyle\cdot\left(\frac{\XCI}{3\times10^{-5}}\right)^{-1}$ \\
        \vspace{1pt}
        CO                      & (0.73  $\pm$ 0.09)$\displaystyle\cdot\left(\frac{r_{41}}{0.46}\right)^{-1}\left(\frac{\alphaCO}{0.8}\right)$ \\
        \vspace{1pt}
        Dust continuum (Band 6) & (2.2 $\pm$ 0.2)$\displaystyle\cdot\left(\frac{\alphadust}{6.7\times10^{19}}\right)^{-1}$ \\
        \hline
    \end{tabular}
    \end{center}
    Table Note: 
    \XCI\ is the atomic carbon abundance, $r_{41}$\ is the ratio of \COfourthree\ to \COonezero\ line luminosity, $\alphaCO$=\mmol$/L^{'}_{\text{\CItwoone}}$\ is the CO conversion factor in the unit of \msun/(K$\cdot$\kms$\cdot$pc$^2$), $\alphadust=L_{\text{dust,850\um}}/\text{\mmol}$ is the dust-to-gas ratio in the unit of erg\,s$^{-1}$\,Hz$^{-1}$ \citep[see][for the derivation and a discussion on its value]{Scoville2016}.
    The quoted uncertainties only include measurement errors.
\end{table}

\subsubsection{Uncertainties of the inferred molecular gas mass}\label{sec:mol_mass_sys_error}
The molecular gas mass inferred using three different tracers are all subject to systematic uncertainties.
For the atomic carbon approach, the excitation of atomic carbon may not be well-constrained from the line ratio alone due to the potential sub-thermal excitation in the diffuse gas \citep{Papadopoulos2022}. The abundance of atomic carbon, \XCI, which depends on the metallicity,
further affects the conversion from \mCI\ to \mmol.
For the CO approach, systematic uncertainties are inherent in the assumed excitation condition, $r_{41}$ \citep{Narayanan2014}, and the $\alphaCO$ factor, the latter having a metallicity dependence as does atomic carbon \citep{Bolatto2013}.
Moreover, the use of the line ratio to infer the gas temperature in the atomic carbon approach, and the reliance on an assumed $r_{41}$ in the CO approach, both require the implicit assumption that the measured line fluxes are representative of the entire galaxy and that the line ratios are computed as globally averaged quantities.
In reality, different line fluxes may be the most prominent in different spatial regions with different physical conditions.
The computation of line ratios across different regions may thus introduce further systematic errors.
For the dust emission approach, the systematic error arises from the assumed dust temperature, emissivity, and spectral index, $\alpha = 1.8$, used to infer the strength of dust emission at $850\,$\um.
Those effects manifest themselves collectively in the error of conversion factor $\alphadust$, which according to the calibration provided by \citet{Scoville2016} is 26 percent.

Amongst the three different estimates of \mmol\ (Table~\ref{tab:mmol_mass}), the one inferred from CO is the most discrepant from that of \CI\ and dust, being lower by a factor of $\approx4$ and $3$, respectively.
This discrepancy can be mitigated by having a higher $\alphaCO$ value or lower $r_{41}$ in the $L_{\rm CO(4-3)}$ to \mmol\ conversion. 
\citet{Dunne2022} conducted a cross-calibration of the three \mmol\ tracers (i.e CO (1-0), \CIonezero, and dust continuum) using a sample of 407 metal-rich galaxies including SMGs like \src. Contrary to previous studies, they report that a universal Milky-Way like $\alphaCO$ value of 4.0 is adequate and there is no evidence for a bimodal $\alphaCO$ as claimed in many previous studies. 
Indeed, in \src\ adopting an $\alphaCO$=4.0 will increase the \mmol\ by a factor of 5, bringing the inferred \mmol\ to be in agreement with \CI\ and dust estimates.
On the other hand, adopting a Milky Way value $r_{41}=0.17$ \citep{Carilli2013} instead of the SMG-value will bring the inferred \mmol\ estimate up by a factor of $\approx2.7$ closer to the other two values.
Some high-redshift dusty star-forming galaxies are found to have higher observed \COonezero\ fluxes that than accounted for in the large velocity gradient modeling that uses high-$J$ CO lines alone, which could indicate the presence of a reservoir of diffuse, low-excitation gas \citep{Canameras2018}.
Therefore, a potential cold gas component with little high-$J$ CO excitation may also explain the \mmol\ discrepancies in our \CI\ and CO approaches.

An alternative explanation for the molecular gas mass discrepancy is that the values inferred using \CI\ and dust are both over-estimated relative to that of CO. Indeed, using the same ALMA observations of \CItwoone\ presented in this work, Lin et al. (accepted in \aap) reported a dynamical mass of $\sim10^{11}$\,\msun\ within the central $3.5$\,kpc in \src\ (see also Sec.~\ref{sec:discussion}).
The total baryonic mass budget, including stellar and gas masses,  exceeds the dynamical mass. The lower molecular mass derived using CO may therefore be more appropriate.
Adopting a Milky Way value of \XCI$=1.6\times10^{-5}$ as reported in \citet{Dunne2022} would bring the \mmol\ inferred from \CI\ up by a factor of $\approx1.9$, which increases the tension with the dust-inferred \mmol.
Alleviating the tension would require hotter dust than the currently assumed $\Td=25$\,K and hence higher $\Gamma_{\rm RJ}$ and/or lower $\alphadust$.
To properly constrain the integrated dust properties, high-resolution continuum observations across frequencies are needed to improve the SED fitting done in \citet{Falkendal2019}.

\subsection{Line ratios}

The ratio between the \CI\ and CO lines can be used to infer the molecular gas excitation conditions.
In particular, the \CItwoone\ to \CIonezero\ luminosity ratio probes the gas excitation temperature.

\begin{figure*}
	\includegraphics[width=\textwidth]{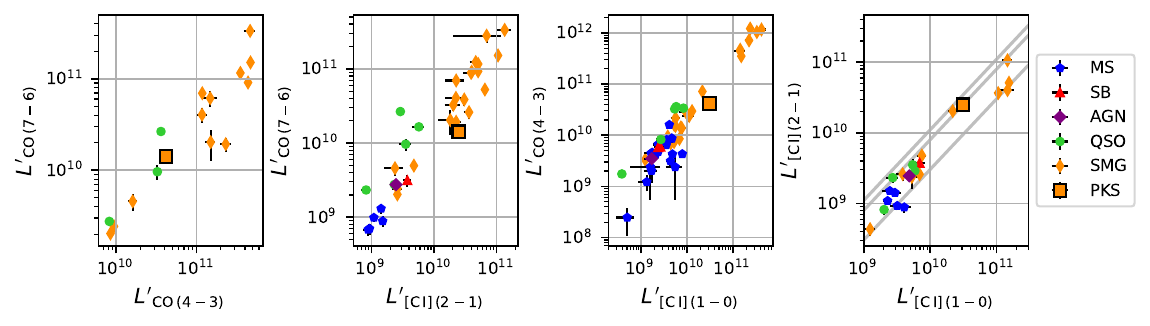}
    \caption{The comparison of various emission line luminosities with the high-z galaxy sample compiled by \citet{Valentino2020}.
    The line luminosities are presented in units of K$\cdot$\kms$\cdot$pc$^2$.
    The classification of galaxies follows that of \citet{Valentino2020}: MS = main sequence; SB = starburst; AGN/QSO = active galactic nucleus/quasar; SMG = submillimeter galaxy.
    Our target, \src, is indicated as an orange box with black edges in the figure.
    For the rightmost panel, $\LineLum{\CItwoone}$\ vs. $\LineLum{\CIonezero}$, we added the lines that indicate the gas temperatures of 20, 40, 60 K (from bottom to top), using the prescription given in \citet{Schneider2003}\ and assuming that \CI\ emissions are optically thin and in LTE condition.
    }
    \label{fig:emission_line_compare}
\end{figure*}

We compare these spatially-integrated line ratios of \src\ with those of the compilation of high-redshift galaxies presented in \citet{Valentino2020},
which include main-sequence, starburst, active galactic nucleus/quasar, and SMGs (see their Sec.~2 for a detailed description of their sample selection).
The comparison of the line ratios are shown in Fig.~\ref{fig:emission_line_compare}.
Due to the heterogeneity in the galaxy types and observations in their compilation, a galaxy is only plotted if it has a detection in both quantities of the axes in each panel.
Overall, \src\ has relatively high emission line luminosities, as expected for its high SFR, similar to the SMG sample.
The only exception is that its $\LineLum{\COsevensix}$ and $\LineLum{\COfourthree}$ are generally at the faint end of the SMG sample, similar to the three QSOs shown in the leftmost panel of Fig.~\ref{fig:emission_line_compare}.
The \COsevensix/\COfourthree\ ratio of \src\ is $r_{74}=0.34\pm0.06$, 
which are comparable to the value reported in \citet[$r_{74}=0.44\pm0.12$]{Bothwell2013a},
slightly higher than that of SMM\,J2135-0102 \citep[][``The Cosmic Eyelash''; $r_{74}=0.24\pm0.03$]{Danielson2011},
and lower than the compilation of SMGs in \citet[$r_{74}=0.76\pm0.17$]{Birkin2021}.
In the rightmost panel, $\LineLum{\CItwoone}$\ vs. $\LineLum{\CIonezero}$, we overlay the diagonal lines that indicate the corresponding gas temperature under the LTE condition using the prescription in Section \ref{sec:mol_mass_CI} \citep[see also][]{Schneider2003,Papadopoulos2004a}.
The gas excitation temperature of \src, $\Tex=44\,\pm\,16$\,K is higher than the median of the SMG samples \citep[$\Tex\approx24$\,K]{Valentino2020}, and more similar to that of QSOs (though see \citealt{Papadopoulos2022} on the possibility of subthermal \CI\ excitation affecting $\Tex$ derivation in the \citealt{Valentino2020} compilation).
The derived carbon excitation temperature is comparable to the poorly constrained dust temperature of $\Td=45^{+23}_{-17}\,$K derived from the SED fitting \citep[western component;][]{Falkendal2019}.
Considering the \CI\ line ratio and the $\Tex$-to-$\Td$ ratio, there does not appear to be a case for subthermal excitation of the \CI\ lines in this source \citep{Harrington2021,Papadopoulos2022}.

\section{Discussion}\label{sec:discussion}

In the new ALMA observations we detected the emission lines of molecular gas tracers, \CItwoone, \CIonezero, \COsevensix, \COfourthree, and continuum emission, from the HzRG \src.
This enables us to derive the molecular gas mass using three different approaches as described in Section \ref{sec:mol_mass}.
The results are presented in Table \ref{tab:mmol_mass}.
In addition to obtaining the integrated spectrum, the high angular resolution ALMA observations allowed us to spatially resolve the emission lines and the continuum.
Dynamical modeling of the \CItwoone\ emission indicates that the molecular gas is a compact and dynamically cold disk supported by rotation (\citealt{Lelli2018}, Lin et al. accepted in \aap), misaligned with the more extended ionized gas emission that trace a large-scale AGN-driven outflow \citep{Nesvadba2017a,Nesvadba2017b}.

The increased sensitivity in the Cycle 6 observations enable us to detect lower surface brightness emission beyond the host galaxy of \src, with the west and east edges overlapping with lobes of radio emission (Fig.~\ref{fig:moment0_overview}).
This suggests that the radio jets may have just broken out of the molecular ISM and not yet of the ionized ISM of the galaxy.

The emissions of \CItwoone\ and \COsevensix\ in \src\ show extended components in the Southwest direction with unclear origins. As noted in Lin et al. (accepted in \aap), there is also a second, weaker \CItwoone\, tail to the East and a kinematically anomalous component near the galaxy center. A possible interpretation is that \src\, is the result of a past, gas-rich, major merger event that may have triggered the starburst and AGN activity.
Radio galaxies have been shown to preferentially reside in overdense environments and possibly in protoclusters \citep{Miley2008,Wylezalek2013, Mei2023}, where galaxy interactions and mergers are expected to be frequent \citep{Lotz2013}.
The observed extended emission in \src\ may also be connected to more diffuse structures on larger scales, which may have been missed by existing ALMA observations.
One of the most prominent examples of a large-scale molecular gas emission surrounding radio galaxies is the HzRG Spiderweb, which in addition to the large-scale molecular gas observed by Very Large Array \citep[70 kpc,][]{Emonts2016}, is a confirmed protocluster with many galaxy companions \citep{Emonts2018,Jin2021}.
Recently, HzRG 4C 41.17 ($z\approx3.8$) has been discovered to have giant filaments of cold gas with sizes up to 100 kpc, using ALMA observations of the \CIonezero\ line \citep{Emonts2023}.
From analyzing the distribution and velocity of the \CIonezero\ emission, the filamentary gas was interpreted as the accretion stream feeding the central radio galaxy.
Other examples of large-scale molecular gas surrounding HzRG include TXS 0828+193 \citep{Fogasy2021}\ and 4C 19.71 \citep{Falkendal2021}.
Regardless of the nature of the large-scale structures found in HzRGs (circumgalactic medium, filamentary gas, or satellite galaxies), their discovery requires observations sensitive to spatially extended emission, i.e. compact configurations of interferometers or single-dish telescopes with large collecting areas such as AtLAST \citep{Lee2024}.

Intriguingly, Lin et al. (accepted in \aap) reported that the baryonic mass of \src\  exceeds its dynamical mass within the innermost $\approx3.5$\,kpc, motivating the need to revisit the assumptions involved in the mass budget.
More accurate stellar mass profiles can be obtained with high-resolution near-infrared imaging with the \textit{James Webb Space Telescope}.
For improving the \mmol\ estimate, observations of lower-$J$ CO lines is essential for normalizing the SLED and deriving \mmol\ with minimal excitation correction.
High-resolution and deeper ALMA Band 4 observations of the \CIonezero\ and CO (4-3) lines will enable spatially-resolved spectral line ratio modeling, in order to characterize the gas temperature and density within and around the host galaxy of \src.
For example, the \CItwoone/\CIonezero\ ratio is determined by the  excitation temperature, the variation in cosmic ray energy density is reflected in the \CI/CO ratio \citep{Bisbas2015}, while the dust-to-gas ratio depends on metallicity \citep{Draine2007}.
We defer this spatially-resolved analysis to a future work when deeper ALMA Band 4 observations becomes available to quantify the gas excitation conditions.

\section{Conclusions}

Deep ALMA observations of \src\ allow us to gain a clearer understanding of the molecular gas in a highly star-forming radio galaxy at $z=2.5725$.
In total, four emission lines, \CItwoone, \COsevensix\ in Band 6, and \CIonezero, \COfourthree\ in Band 4, were detected and spatially resolved (see Figure \ref{fig:moment0_overview} for the line intensity maps and Figure \ref{fig:spectrum_overview} for the integrated spectra).
Out of the four emission lines, \CItwoone\ was already previously detected in a shallower ALMA observation \citep{Man2019} and used to dynamically model the galaxy \citep{Lelli2018}.
In addition to the rotating disk inferred from the previous ALMA observation \citep{Lelli2018}, the new, deeper ALMA observations show that \CItwoone\ has a diffuse and extended component that was missed previously due to limited sensitivity (see Lin et al. accepted in \aap for the accompanying kinematic modeling analysis).

The continuum images of \src\ in both Bands detect the eastern radio lobe and the central galaxy \citep[Fig.~\ref{fig:cont_fit}; ][]{Broderick2007,Falkendal2019}.
In Band 4, the additional synchrotron component of the western radio lobe is detected.

The detection of \CI, CO lines and dust continuum emission enables us to use three different approaches to infer the mass of molecular gas \mmol\ in \src\ (Sec.~\ref{sec:mol_mass}).
The three approaches indicate that \mmol\ is $\approx1-3\times10^{11}$\msun) (Table~\ref{tab:mmol_mass}).
The \mmol\ inferred from \CI\ and dust are in good agreement,
while the CO-based \mmol\ is a factor of $3-4$ lower though can be brought to consistency with the other two estimates if we adopt Milky Way values for the CO excitation or the $\alphaCO$ as advocated for in \citet{Dunne2022}.
Regardless, all three approaches indicate nearly an order of magnitude higher molecular gas reservoir than the previous estimate \citep[][see Appendix~\ref{sec:CI_comparison}]{Man2019}.

Lastly, the extended \CItwoone\ and \COsevensix\ emission in the southeastern direction may be related to larger structures as discovered in other HzRGs \citep[such as tidal tails from a
past major merger,
see Lin et al. accepted in \aap; ][]{Emonts2015a,Emonts2015b,Falkendal2021,Fogasy2021,Emonts2023}.
Deeper observations covering larger spatial scales are needed to uncover the nature of this extended emission.

\begin{acknowledgments}
We are grateful to Padelis Papadopoulos, Paola Andreani, Matthew Lehnert, and Kevin Harrington for constructive comments that improved this manuscript.
HTH acknowledges the support of the Summer Undergraduate Research Exchange programme at the Chinese University of Hong Kong.
AWSM and LG acknowledge the support of the Natural Sciences and Engineering Research Council of Canada (NSERC) through grant reference number RGPIN-2021-03046.
AWSM acknowledges the support through the ESO Visitor Programme that facilitated the completion of this work.
ZYZ, LL and JZ acknowledge the support from the National Key R\&D Program of China (2023YFA1608204), the National Natural Science Foundation of China (NSFC) under grants 12173016 and 12041305, the science research grants from the China Manned Space Project, CMS-CSST-2021-A08 and CMS-CSST-2021-A07, and the Program for Innovative Talents, Entrepreneur in Jiangsu. 
TGB acknowledges support from the Leading Innovation and Entrepreneurship Team of Zhejiang Province of China (Grant No. 2023R01008).
This paper makes use of data from ALMA program ADS/JAO.ALMA\#2013.1.00521.S, ADS/JAO.ALMA\#2018.1.01669.S. 
ALMA is a partnership of ESO (representing its member states), NSF (USA) and NINS (Japan), together
with NRC (Canada) and NSC and ASIAA (Taiwan) and KASI (Republic of Korea),
in cooperation with the Republic of Chile. 
The Joint ALMA Observatory is operated by ESO, AUI/NRAO and NAOJ.
\end{acknowledgments}

\facility{ALMA}

\appendix
\section{Comparison with previous molecular gas mass measurement of \src\ using \CItwoone}\label{sec:CI_comparison}
The mass of molecular gas we derived above using \CItwoone\ and \CIonezero\ is about an order of magnitude larger than the value obtained in \citet{Man2019} using only \CItwoone\ flux in the shallower ALMA Cycle 2 observations ($(3.1\pm0.6)\times10^{11}$\msun\ versus $(3.9\pm1.0)\times10^{10}$\msun).\footnote{The value in \citet{Man2019}, $(3.9\pm1.0)\times10^{10}$\msun, also includes the 1.36 correction factor for the Helium and heavier elements.}
This difference can be attributed to the combination of three factors, each one giving a factor of about 2 difference in the final molecular mass:
\begin{enumerate}
    \item Deeper Cycle 6 ALMA observation used in this work allows us to discover a more spatially extended emission of \CItwoone\ that was previously undetected in the shallower Cycle 2 observations (see Figure~\ref{fig:moment0_overview}).
    This newly determined continuum level is also slightly lower than that determined in the shallower Cycle 2 observations presented in \citet{Man2019}. 
    The combined effect of a larger emission line region (and thus spectrum extraction aperture) and a lowered continuum level increases the integrated flux of \CItwoone\ by a factor of 2 (4.04 $\pm$ 0.30 versus 2.0 $\pm$ 0.5 Jy$\cdot$\kms). If the same aperture in \citet{Man2019} were applied to our new ALMA observation data, it would miss the extended flux and get the integrated flux of \CItwoone\ 2.0 $\pm$ 0.1 Jy$\cdot$\kms.
    \item In \citet{Man2019}\ only \CItwoone\ was observed and the excitation condition of \CI\ was assumed in order to derive the total molecular gas mass. 
    The factor $Q_{21}$ is defined as the ratio of atomic carbon in $^3$P$_{2}$\ state to the entire population \citep[][see also footnote \ref{footnote:Q}]{Papadopoulos2004a}. Without \CIonezero\ observations, \citet{Man2019}\ assumed $Q_{21}$\ to be 0.5.
    In this work, we can directly infer $\Tex$\ using the flux ratio of the \CItwoone\ and \CIonezero\ lines (under the assumption of LTE condition), which gives $Q_{21}=0.29\,\pm\,0.06$. This results in a factor of 1.7 difference in the molecular gas mass estimation.
    \item In the derivation of molecular gas mass, \citet{Man2019}\ used equation (12) from \citet{Papadopoulos2004a}\ to convert the flux of \CItwoone\ to \mmol. 
    However, this equation implicitly calculates the luminosity distance from the cosmology without dark energy.
    The exclusion of dark energy caused the luminosity distance to be underestimated compared to the cosmology in which $\Omega_\mathrm{M}$\ = 0.3 and $\Omega_{\Lambda}$\ = 0.7.
    This in turn resulted in the underestimation of \CItwoone\ line luminosity in \citet{Man2019} by a factor of 2.2.
\end{enumerate}

\bibliography{pks}{}
\bibliographystyle{aasjournal}

\end{document}